\documentclass[fleqn]{emulateapj}
\usepackage[T1]{fontenc}
\usepackage{ae,aecompl}
\usepackage{graphicx}	% Including figure files

\usepackage{ulem}
\usepackage{amsmath}	% Advanced maths commands
\usepackage{amssymb}	% Extra maths symbols
\usepackage{color}      
\newcommand{\sn}{iPTF16geu}
\newcommand{\simgt}{\hbox{\,\rlap{\raise 0.425ex\hbox{$>$}}\lower 0.65ex\hbox{$\sim$}\,}}
\newcommand{\simlt}{\hbox{\,\rlap{\raise 0.425ex\hbox{$<$}}\lower 0.65ex\hbox{$\sim$}\,}}

\slugcomment{Submitted to The Open Journal of Astrophysics}

\shorttitle{Lensing galaxy in iPTF16geu}
\shortauthors{Williams \& Zegeye}

\begin{document}

\title{Two-component mass models of the lensing galaxy in the quadruply imaged supernova \sn}

\author{Liliya L.R. Williams}
\affil{School of Physics and Astronomy, University of Minnesota, 116 Church Street, Minneapolis, MN 55455, USA}
\email{llrw@umn.edu (LLRW)}

\and

\author{David Zegeye}
\affil{Department of Astronomy and Astrophysics, University of Chicago, 5640 South Ellis Avenue, Chicago, IL 60637, USA}

\begin{abstract}
The first resolved, multiply imaged supernova Type Ia, \sn, was observed 4 years ago, five decades after such systems were first envisioned. Because of the unique properties of the source, these systems hold a lot of promise for the study of galaxy structure and cosmological parameters. However, this very first example presented modelers with a few puzzles. It was expected that to explain image fluxes a contribution from microlensing by stars would be required, but to accommodate the magnitude of microlensing, the density slope of the elliptical power law lens model had to be quite shallow, $\rho_{2D}\propto r^{-0.7}$. Furthermore, the center of mass had to be displaced from that of observed light by $\sim 0.1$kpc, and the position angle of light distribution was misaligned with that of mass by $\sim 40^o$. In this paper we present mass models that resolve the first two problems, and suggest a resolution of the third. Motivated by observations of local ellipticals, and some recent analysis of galaxy-scale lenses, our mass models consist of two offset (baryonic) mass components. The resulting mass distributions have a single centroid, but are lopsided, and have isodensity contours that are not purely elliptical and not self-similar with radius. For many of our models the microlensing requirements are modest, and the ring formed by the extended supernova host galaxy resembles the observed one.
\end{abstract}


\section{Introduction}   

The pioneering work of \cite{ref64} described a new method to measure the Hubble constant, using a phenomenon---gravitational lensing---that would not be observed for another 15 years \citep{wal79}. Though the lensed source was envisioned to be a supernova, multiply imaged supernova were not detected until only a few years ago \citep{qui14,kel15,goo17}. 

There is considerable interest in multiply imaged supernova, and especially of Type Ia, because of the wealth of information that can be extracted from them \citep[e.g.,][]{ogu19,wag19}, including mass distribution in galaxies \citep{ogu03,jon10}, global cosmology \citep{qi19}, and constraints on the nature of dark matter \citep{zum18,pan14}. Systems with lensed supernova present some challenges as well, for example, finding these in surveys \citep{woj19,por00}, and accounting for microlensing of the multiple images \citep{dob06,yah17,fox18,pie19,gol18}.

Supernova \sn~was the first Type Ia supernova with resolved multiple images \citep{goo17}. The source and lens are at redshifts $z=0.409$ and $z=0.2163$, respectively. Using supernova lightcurve fitting, \cite{dha20} estimate that the total magnification is $\mu=67.8\pm 2.7$ at 68\% confidence level. Microlensing by stars in the lensing galaxy was recognized to be important in this system since it was first detected \citep{fox18,mor17}. However, it was soon realized that the combination of proposed macro-lensing (i.e., due to the main lensing galaxy) and microlensing models are not able to fully explain the system. Specifically, assuming that the galaxy lens is described by a power law projected density profile with an ``isothermal" slope, $\Sigma\propto r^{-1}$, would require a considerable, even unrealistic amount of microlensing (de)magnification of the images \citep{yah17}.  Varying the fraction of dark matter compared to total mass, or including external shear does not help \citep{yah17}.

This conclusion suggests that an isothermal power law lens is unlikely. \cite{mor19} considered a range of density profile slopes, $\Sigma\propto r^{\alpha-2}$, and conclude that a shallower galaxy lens will result in predicted macro-magnifications that will not put too much strain on microlensing contribution. The best fitting slope from microlensing considerations is $\alpha\approx 1.3$, significantly shallower than isothermal ($\alpha=1$). Such shallow slopes are far from SLACS findings \citep{koo06}, of $\alpha=0.99\pm 0.12$, but in the range of profile slopes found in \cite{rea07} for lensing galaxies.

While the central regions of some galaxies may have shallow density profiles, the profiles must steepen at larger radii. In addition to this, a further feature of their model suggests that it might be a simplification of the actual lens: the center of observed light is displaced from the center of recovered mass by $\sim 0.03''$, or about  $0.1$kpc, much larger than the astrometric uncertainties of $0.002''$. Finally, the position angle of the fitted elliptical mass distribution is misaligned with the observed light orientation by $\sim 40^\circ$. 

Taken together, these properties of the lens model lead us to conclude that the actual mass distribution is more complicated than an elliptical density power law. What form could these complications take?  Addition of dark matter substructure, as predicted by $\Lambda$CDM cosmological model \citep{kly99,moo99} is unlikely to reconcile the simple mass model with observations because substructure cannot change profile slope, or affect image positions significantly enough to reconcile light and mass centroids. Microlensing by stars can significantly affect image fluxes, but not image positions. The additional complications must be due to the macro-scale mass model of the galaxy lens. 

\cite{wag18} has shown that images provide only local information about the lensing potential; the rest is usually filled in by model assumptions. This motivates us to explore other possibilities for what projected lens looks like in the case of \sn.  Specifically, we extend the published lens models beyond a single power law; our models consist of two superimposed, cored power law mass components, each with its own set of parameters (\S~\ref{galmod}). Our recovered mass distributions, generated using the method described in \S~\ref{modfit}, account for image positions, and produce coincident light and mass centroids within astrometric error (\S~\ref{res}). We do not use time delays, because of the large uncertainties, or the extended image of the supernova host galaxy, which was found by \cite{mor19} not to be very useful. We do compare extended rings and time delays generated by our models with observations, in \S~\ref{ring} and \S~\ref{td}.

\section{Galaxy model}\label{galmod}

To motivate our mass model we assume that the lens galaxy at $z\approx0.2$ must be analogous to nearby ellipticals of similar dynamical mass and effective radius of the light distribution, $R_e$. Its velocity dispersion is 129 km/s \citep{joh20}\footnote{An earlier estimate by \cite{mor19} was 164 km/s.}, and  \cite{dha20} measure $R_e\approx 0.55''$ (depending on the filter), or 1.83 kpc, at the redshift of the lens. 

Virgo cluster ellipticals are well studied, and some are analogues to the \sn~lensing galaxy. For example, NGC 4494, has $\sigma\sim 150$ km/s \citep{rom03}, and $R_e\sim 3.8$ kpc \citep{fos11}, and NGC 4434 has $\sigma\sim 100$ km/s \citep{dre84}, and $R_e\sim 1.5$ kpc \citep{kor09,kra18}.  Central regions of galaxies inside the light's effective radius are dominated by stars. This is especially true of lower mass ellipticals \citep{fer05}. In \sn, quad images are formed at $\sim 1$kpc, or $\sim 0.55 R_e$, where the bulk of the mass is in the form of stars. Therefore, to motivate our mass model we need to look at the stellar mass distribution in low mass ellipticals. 

\cite{dha12} used data from \cite{kor09} to fit sky projected radial light distribution of Virgo ellipticals over $4-5$ decades in radius. A single component, either a power law or a Sersic profile does not provide a good fit, while fits using 2 or 3 projected Einasto components have very small residuals, with rms of $0.025-0.065$ magnitudes for NGC 4494 and 4434. When fitted with 2 components, the ratio of the two scale-lengths is about $10-25$, for both of these galaxies. If these Virgo galaxies were the lensing galaxies in \sn, images would form somewhat interior to the scale-length of the larger component, at the radius where the total light distribution undergoes the most rapid change in (projected) slope. 

Our mass models represent the lensing galaxy as a superposition of two mass components, but instead of using projected Einasto profiles, which do not have analytical expressions for lensing potential and deflection angles, we use {\tt alphapot} analytical lensing potential \citep{kee11}, whose expressions for deflection angles and projected density, or convergence are presented, for example, in \cite{gho20}. Each {\tt alphapot} is characterized by a center, normalization $b$, potential profile slope $\alpha$, core radius $s$, and ellipticity parameters $q$, and $K$:
\begin{equation}
\Psi=b(s^2+x^2+y^2/q^2+K^2xy)^{\alpha/2}
\label{alphapot}
\end{equation}
If $K=0$, the elongation of the iso-potential contours is along the $x$ or $y$ axis, and $q$ gives the axis ratio. If $K\ne0$, the ellipticity position angle is no longer aligned with the principal axes. Because of adjustable core size and power law density slope, this profile captures the most important feature of Einasto profile over the radial range relevant to lensing, namely the ability of have constant or changing density slope. 

We allow the centers of the two mass components to be offset from each other, which amounts to having a dipole moment in the mass distribution around the center of light. \cite{gom18} showed that to reproduce the statistical properties of image distribution around the lens center of the sample of $\sim 40$ galaxy-scale quad lenses, one needs to introduce small, dipolar-like azimuthal mass perturbations around the lens center. The authors implemented these by offsetting the centers of the two mass components, while the total mass distribution still had a single centroid. Further evidence for such complexity in mass distribution comes from the modeling presented in \cite{nig19}: all three of their massive elliptical lenses consist of two baryonic components, and show lopsidedness in their mass distribution.

The more compact of the two mass components in our models is designated as the main one, while the secondary is more extended. The two need not have equal mass, or brightness. In fact, the central brightness of the secondary component tends to be fainter by about a magnitude \citep{dha12}. In principle, the second component in our models could also represent dark matter, with considerable or minimal stellar contribution. For lensing, all that matters is their mass content.

\begin{figure*}    %% figure 1
\centering
\vspace{-5pt}
\includegraphics[width=0.43\linewidth]{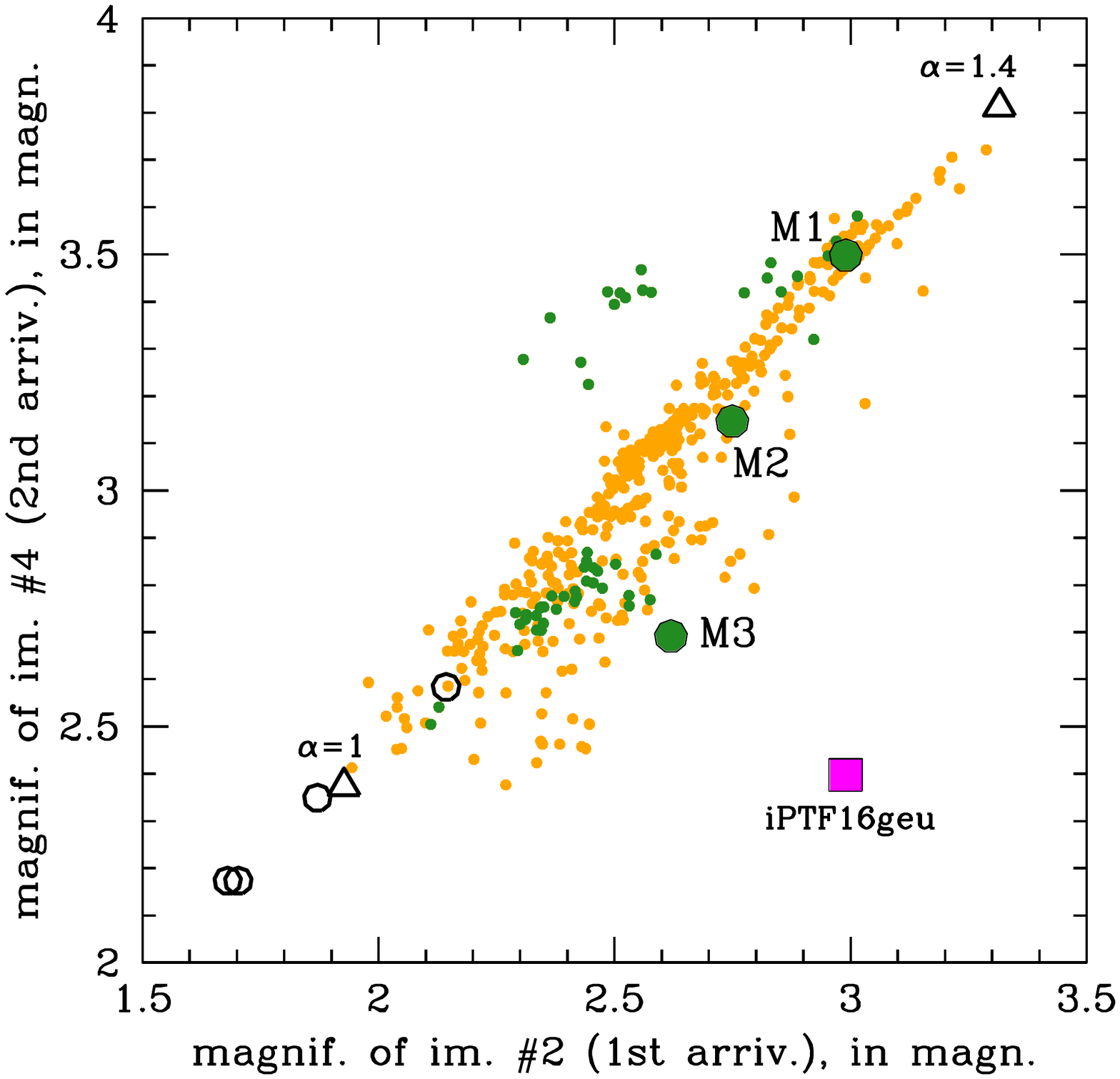}
\includegraphics[width=0.43\linewidth]{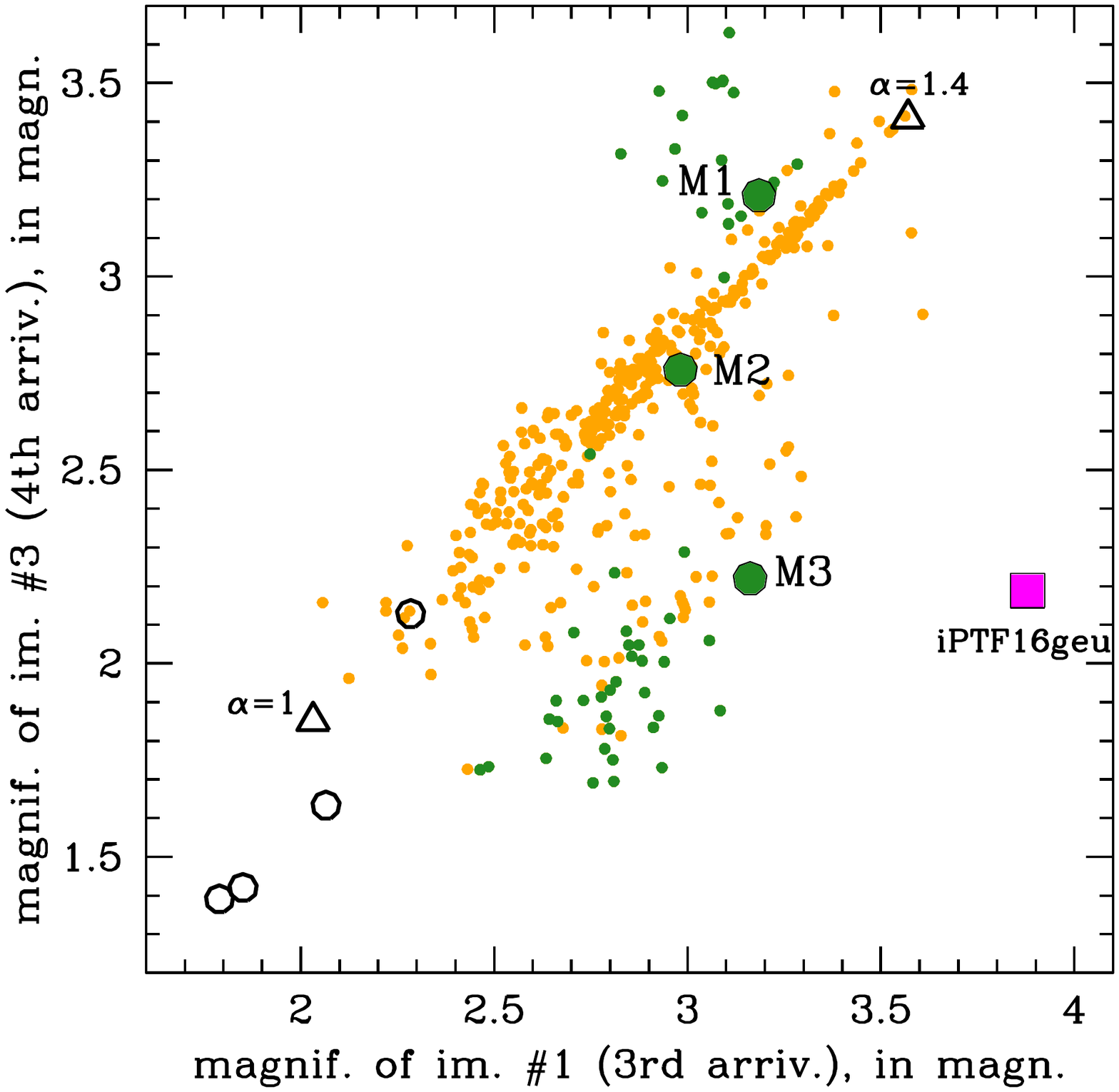}
\vspace{-50pt}
\caption{Image magnifications, in magnitudes. {\it Left panel:} Minima point images: 1st and 2nd arriving (labelled $\#2$ and $\#4$ in previous works). {\it Right panel:} Saddle point images: 3rd and 4th arriving (labelled $\#1$ and $\#3$ in previous works). In both panels, the observed magnifications of \sn~images \protect\citep{dha20} are denoted by a magenta square. Empty black symbols represent published models: circles are \protect\cite{mor17}, and triangles are \protect\cite{mor19} (labeled by the density profile slope). Orange points are our 458 models, all satisfying $\chi^2<1$ for image positions. Green points are a subset of these, chosen based on criteria described at the end of \S~\ref{modfit}.}
\label{smmagnifs}
\end{figure*}

\section{Model fitting}\label{modfit}

The positions of the 4 images of the quad give us 8 data constraints. We do not use the ring formed by the extended light from the supernova host galaxy, as \cite{mor19} found it to be not very useful; the ring did not improve the fit or change if from the one found using point images alone. We do not use time delays either, because they are expected to be short in this cross-type quad, and because of their rather large uncertainties. Later in this Section we present $\chi^2$ values calculated without the time delay contribution, and in \S~\ref{td} we include time delay information, and show that in this system time delays do not help discrimanate between models. We present images of extended host galaxy for some of our final models in \S~\ref{ring} and time delays in \S~\ref{td}.

Our two mass components, $A$ and $B$, have 5 free parameters each, $b$, $s$, $q$, $K$, and $\alpha$, defined in eq.~\eqref{alphapot}. The offset between the two,  $\Delta x, \Delta y$, adds 2 more free parameters, and 2 additional ones are due to the unknown location of the source. From these 14, we eliminate 2 that correspond to the mass normalization of the lens and its orientation (rotation) on the plane of the sky.  In practice that means we set $K_A=0$, and only the ratio $b_B/b_A$ matters, not their individual values. In other words, we find fits to the quad characterized by 3 image distance ratios with respect to the lens center, and 3 relative polar image angles. Having found a fit, the rotation of the whole system with respect to the center of light, and the mass normalization are adjusted to correspond to the images of \sn. Specifically, the model quad system is shrunk or expanded, and rotated, such that the model's first arriving image is exactly at the distance and position angle of the observed first arriving image. 

\begin{table*}
\centering
\begin{tabular}{|c|c|c|c|c|c|c|c|c|c|c|c|}
\hline 
\hline
$\Delta x, y$ & $b_B/b_A$ & $\alpha_A$& $q_A^2$  & $K_A^2$  &$\alpha_B$ & $q_B^2$  & $K_B^2$ \\
\hline 
    $\pm$0.075''   & 1-100 & 0.6--1.4 & 0.7--1.0 & 0& 0.6--1.4 & 0.7--1.3 & -0.4--0.4 \\ 
\hline
\hline
\end{tabular}
\caption{Ranges of parameter values used when randomly searching the 12 dimensional parameter space of the 2 component mass model.}
\label{tableX}
\end{table*}

This means that our merit function is 
\begin{equation}\label{eq:chi2p}
  \chi_p^2=\frac{1}{6}\,\Big(\sum_{i=2}^4\, \Big[\frac{(d_{i,m}-d_{i,o})^2}{(\sigma_{d,i})^2} \Big] 
           + \sum_{i=2}^4\, \Big[\frac{(\theta_{i,m}-\theta_{i,o})^2}{(\sigma_{\theta,i})^2} \Big] \Big),
\end{equation}
where $d_i$ is the ratio of the distance of image $i$ from the lens center to the distance of the first arriving image, and similarly for image angle $\theta_i$. Subscripts $m$ and $o$ refer to the model and observed image properties, respectively.  Uncertainties $\sigma$'s were obtained from the astrometric uncertainties in \cite{mor19}. Subscript $p$ on $\chi^2$ means that it is based on positional data only. In \S~\ref{td} we consider time delays as well, which modifies the merit function to
\begin{equation}\label{eq:chi2pt}
  \chi_{p,t}^2=\frac{1}{9}\,\Big( 6\chi_p^2+ \sum_{j=2}^4\,  \Big[\frac{(t_{j,m}-t_{j,o})^2}{(\sigma_{t,j})^2} \Big]  \Big),
\end{equation}
where the time delays given in \cite{mor19} are with respect to the third arriving image.

We search over a ranges of parameters, given in Table~\ref{tableX}. Ellipticity parameters were not allowed to be too far from circular, in keeping with the appearance of the light of the lensing galaxy. Additionally, the core radii of the two components, $s_A$ and $s_B$, were constrained to be smaller than the image radius.  The ratio of normalization parameters, $b_B/b_A$ was picked from a wide range, while the offset of the second component was confined to about a quarter of the Einstein radius. While these starting parameter ranges are somewhat arbitrary, we note that the second stage of our two-step modeling procedure (see below) allows parameters to wonder well outside of the ranges set in this first step.

We use the following method to search the 12 dimensional parameter space for solutions. Since the model parameters outnumber lensing constraints, we do not expect a unique best fit solution. Instead, many solutions will satisfy the constraints of \sn. Our modeling consists of two stages. First, we randomly search a wide range of parameters, looking for approximate fits to the lens system. It is not hard to find many fits with $\chi_p^2\simlt 9$ using a random search. We generated 300 of these. As expected, the distribution of $\chi_p^2$ values in this sample is heavily biased towards high values. Then, we use randomly selected pairs of these bad solutions to define pairs of points in the 12 dimensional parameter space. The two solutions in a pair serve as ``opposite'' vertices of a simplex. In 2D such a simplex will be a square, in 3D, a cube, etc. Each simplex is input into downhill simplex method to find better solutions. Of the 2000 downhill simplex runs, 458, or about $23\%$ had $\chi_p^2< 1$. 
We checked that these solutions are in fact separate local minima in the 12 dimensional parameter space, because a straight line between any two contains worse fits. Our method, which combines a random search followed by downhill simplex search, is an easily implemented and effective way of finding solutions in multidimensional parameter space that has many local minima. The two steps of our search method resemble those of the particle swarm optimization, though the latter looks for a single solution, while we look for many. Note that there is no need to find the global minimum; all local minima have $\chi_p^2< 1$, and many have $\chi_p^2< 0.1$.  

Because our mass models contain more parameters than data constraints, we need to check that these extra parameters are  statistically justified. To do that,  we compare our models to commonly used,  single mass component models.  To allow for better fits we include external shear, which results in $k=5$ total free parameters.  In our two-component models $k=12$ (see above in this section). When the center of the one-component lens is fixed at the center of the observed light, as it is in the two-component models, a typical $\chi_p^2\sim 27$. Both models have $n=6$ data points (three relative polar image angles and distance ratios, with respect to the lens center). Using Bayesian Information Criterion, BIC~$=\chi^2+k\ln n$, we get $\approx 36$ for one-component models, and $\approx 22.5$ for our two-component models. BIC differences of $>10$ are considered very strong evidence for models with smaller BIC values. Thus, two-component models are well justified in this system.

Each solution is a lens model of \sn, and has associated image magnifications. The two panels of Figure~\ref{smmagnifs} show magnifications of the two images formed at the minima of the arrival time surface (left panel), and two images formed at the saddle points (right panel). We designate the images in two ways: the published papers label them as \#1, ...\#4, and we also add designation based on the arrival time, as 1st, ..., 4th \citep{sah03}.
 
We show these 458 solutions as small orange filled points in these plots. (We checked that the distribution of solutions does not depend on their $\chi_p^2$ values, so a somewhat different value for the $\chi_p^2$ cut would have generated a statistically similar distribution of solutions.)  Most of them lie near a straight line, running approximately diagonally from low to high magnifications in each of the panels. The models from published literature are denoted by empty black triangles \citep[][Table 9]{mor19} and circles \citep[][Table 1]{mor17}, and also lie approximately along the same diagonal lines. Solutions along these two lines share similar geometry of their mass distributions, specifically, they are connected by approximate mass sheet degeneracy (MSD), which is more aptly called the steepness degeneracy \citep{gor88,sah00}; flatter density profiles correspond to higher image magnifications.  Note that MSD here is not exact: rescaling of $\kappa$ of one solution by some $\lambda$ and adding a flat mass sheet of density $1-\lambda$ (in units of critical lensing density) will generate a solution which is only approximately similar to the ones plotted here along these diagonals. The actual relation between these solutions (both orange and empty black symbols) are probably better described by non-trivial source plane transformations \citep[SPT; ][]{sch14}.

Not all of our 458 solutions are equally good representations of galaxies. Because models consist of two offset mass components, many {\it total} mass models will have two separated centroids. These are probably un-astrophysical, and we do not consider them further. In the remaining $\sim 30\%$ of the cases the offset centers merely result in lopsidedness, or dipole-like feature in the mass distribution. The total mass distribution in these has a single centroid; in other words, the center of light and mass coincide.  Aiming to isolate models that are closer to single component ones, we further select solutions where the main (compact) component contains at least half of the total mass within the image ring. We end up with a subset of 57 solutions, plotted as green dots in Figure~\ref{smmagnifs}.   By selecting these 57, we do not preferentially select better or worse fits: we checked that $\chi_p^2$ does not correlate with the fraction of mass in the main component, or   how far the second centroid is from the light center.

Though our imposed cuts do not correlate with the goodness of fit, they do correlate with other lens properties. Most of these 57 solutions deviate from the diagonal lines, especially in the right panel, and many of these lie closer to \sn, than further from it, meaning that these solutions require less microlensing (de)magnification.  

We cannot examine all 57 models in detail. Instead, we pick a small handful of examples,  based on their magnification properties, as these are important for microlensing.  Models M1 and M3 are on opposite sides of the main diagonal in Figure~\ref{smmagnifs}, and are closer to the \sn~than other green points within their respective islands of solutions. Additionally, we selected M3 because it is close to \sn~in both panels of Figure~\ref{smmagnifs}. Thus M3 would require minimal contribution from microlensing. M2 was selected because it is on the main diagonal.  We note that we could have picked a different set of models, or models based on physical criteria other than magnification.  But the wide range of  models allowed by the data requires us to make choices.

\section{Results: examining models M1-M3}\label{res}

\subsection{Mass distributions}

\begin{figure*}    %% figure 2
\centering
\vspace{-5pt}
\includegraphics[width=0.32\linewidth]{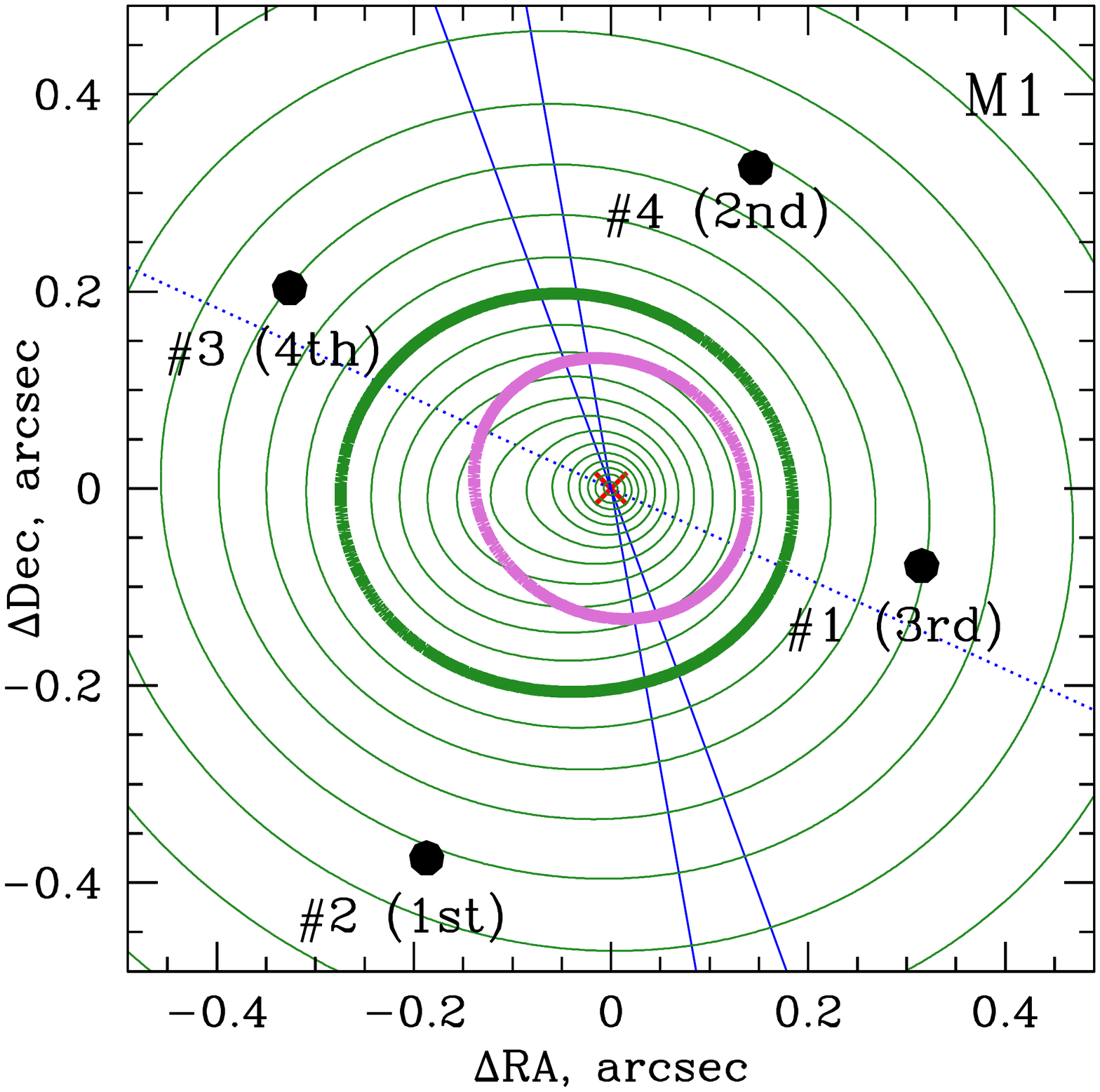}
\includegraphics[width=0.32\linewidth]{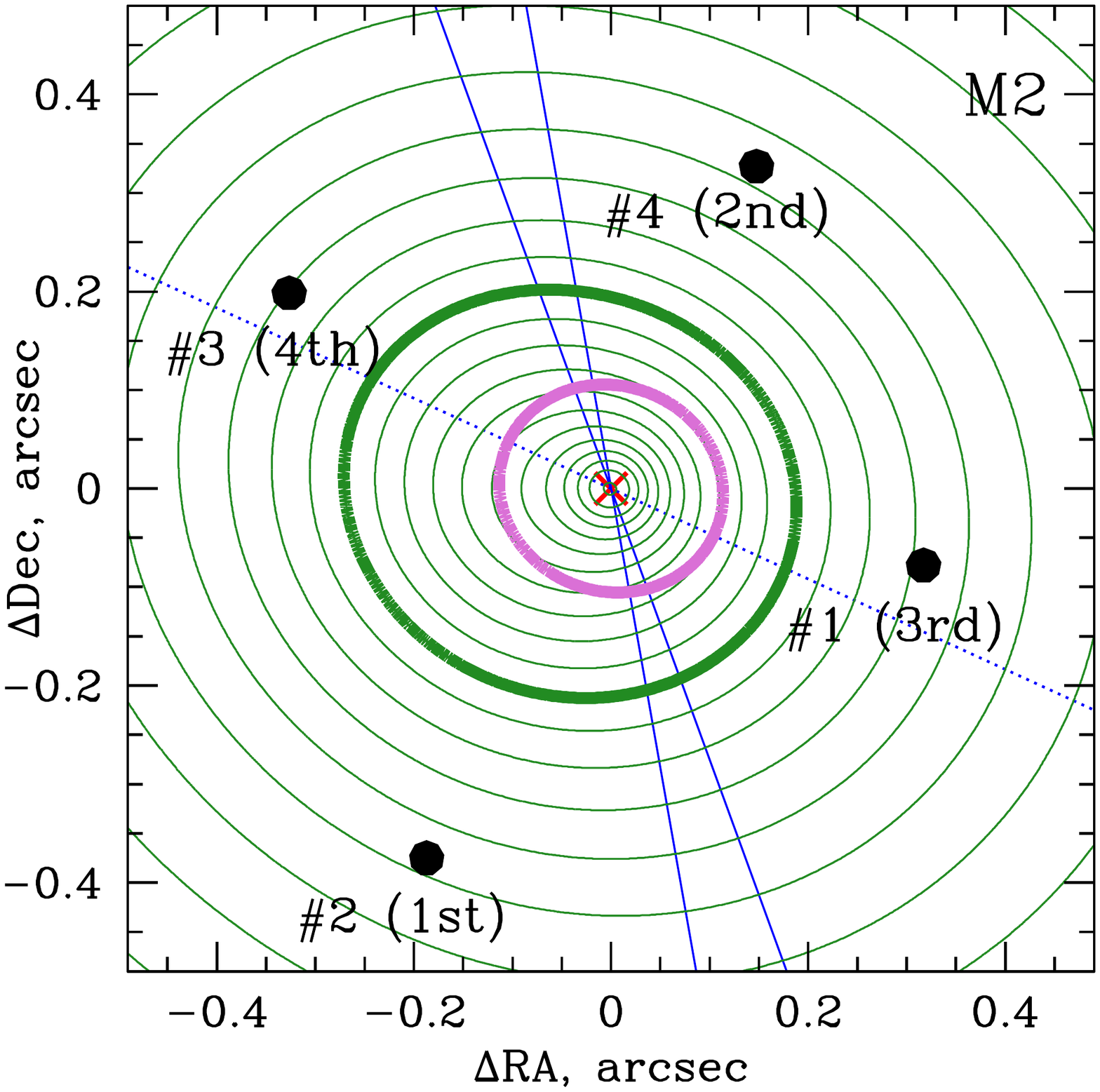}
\includegraphics[width=0.32\linewidth]{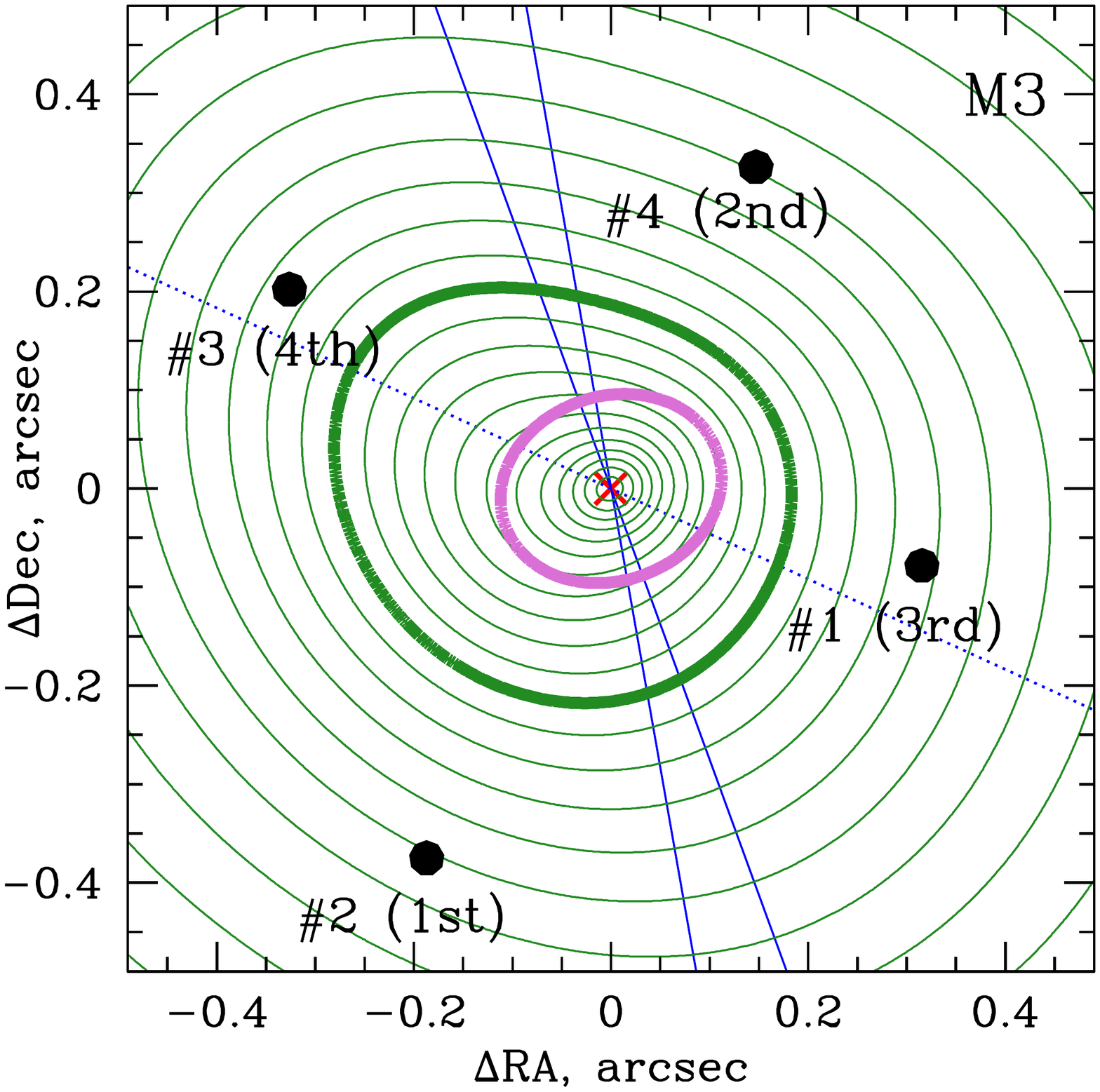}
\vspace{-35pt}
\caption{Three mass models M1-M3, shown in Figure~\ref{smmagnifs}. In all 3, the center of light is coincident with the center of mass, even though the mass contours are not purely elliptical. The thick green and magenta contours represent $\kappa=1$ levels of the total mass and the main (compact) mass component, respectively. The green thin contours of the total mass are spaced by 0.05 in the log. Images are labeled as in \protect\cite{mor19}, and the arrival sequence is given in parentheses. Thin blue line is the orientation of the elliptical mass  models in \protect\cite{mor19}, while the two thicker blue lines give the range of the position angle of the observed light. (North is up and East to the left.)}
\label{smmassrotate}
\end{figure*}

In this section we examine the three solutions selected in the previous section, and depicted as large green filled symbols in Figure~\ref{smmagnifs}. They span a range of magnifications, and have coincident mass and light centroids. We present their mass distributions in Figure~\ref{smmassrotate}.
The magenta is the $\kappa=1$ contour of the main (compact) mass component, while the green curves represent  the isodensity contours of the total mass distribution, logarithmically spaced with an interval of 0.05. The thick green contour denotes $\kappa=1$. The green isodensity contours are not exactly elliptical because they represent a sum of two offset mass components. The fact that the offset results in a dipole-like mass distribution around the lens center is evident in these plots. 

\begin{figure*}    %% figure 3
\centering
\vspace{-5pt}
\includegraphics[width=0.32\linewidth]{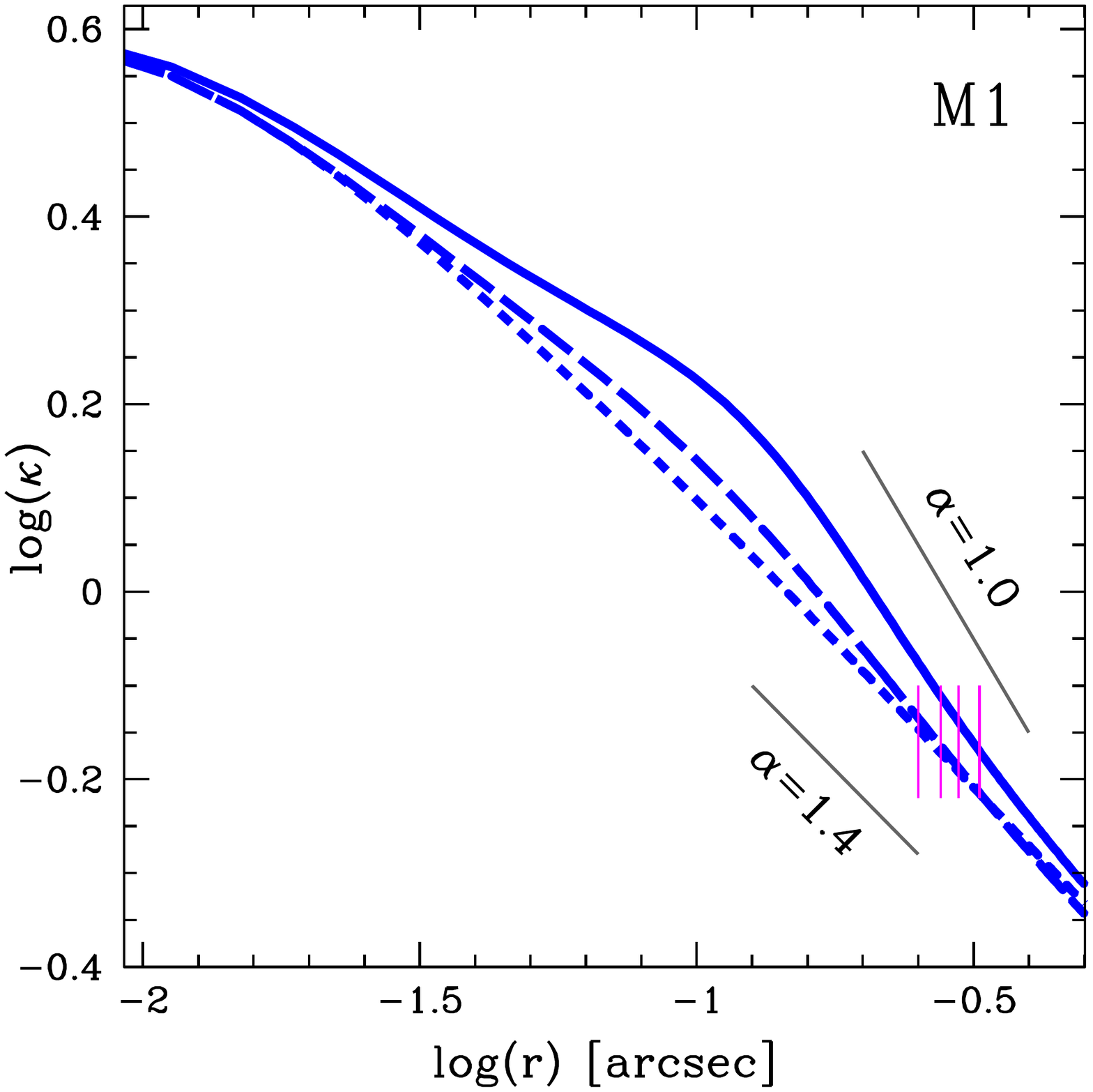}
\includegraphics[width=0.32\linewidth]{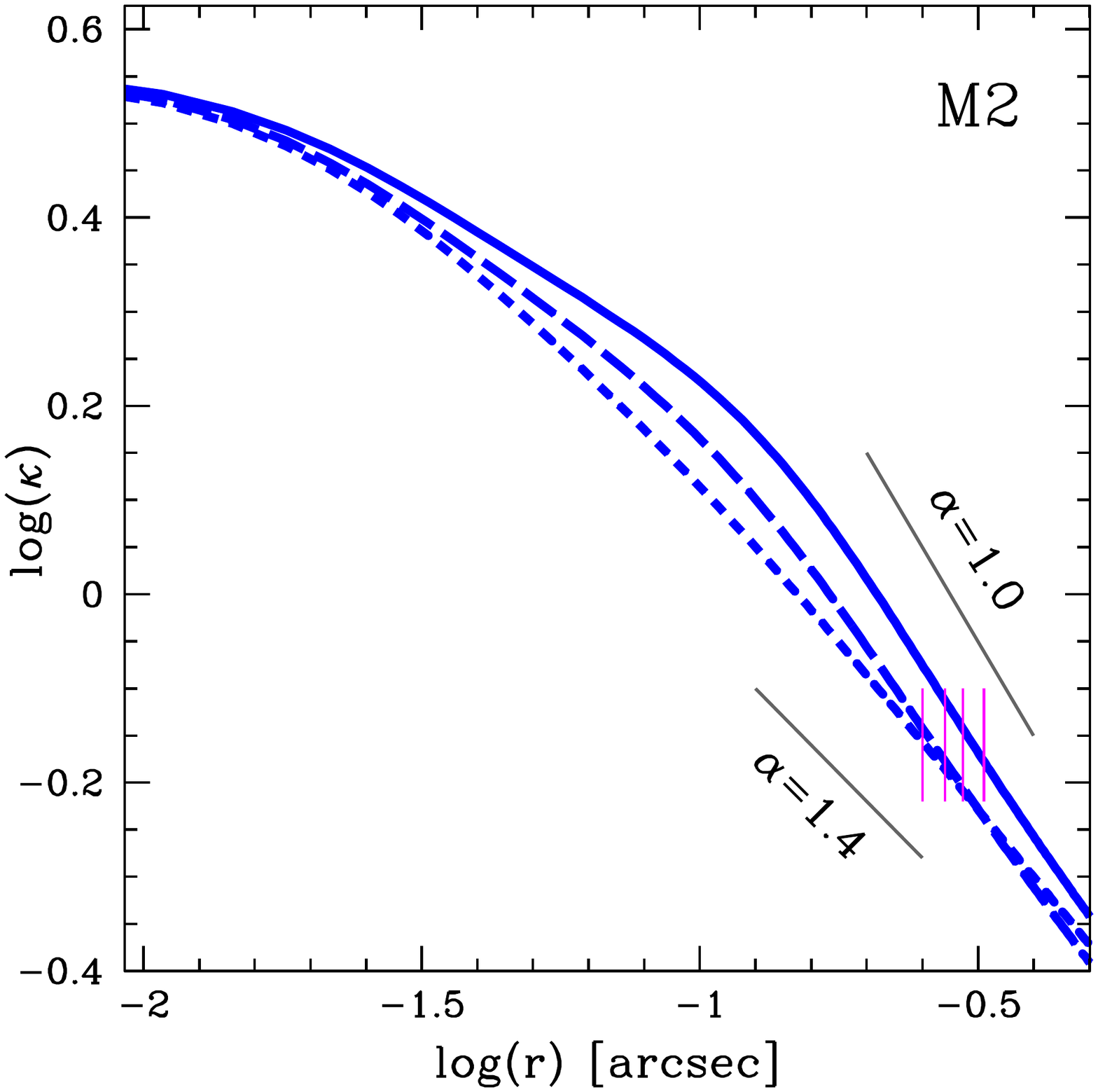}
\includegraphics[width=0.32\linewidth]{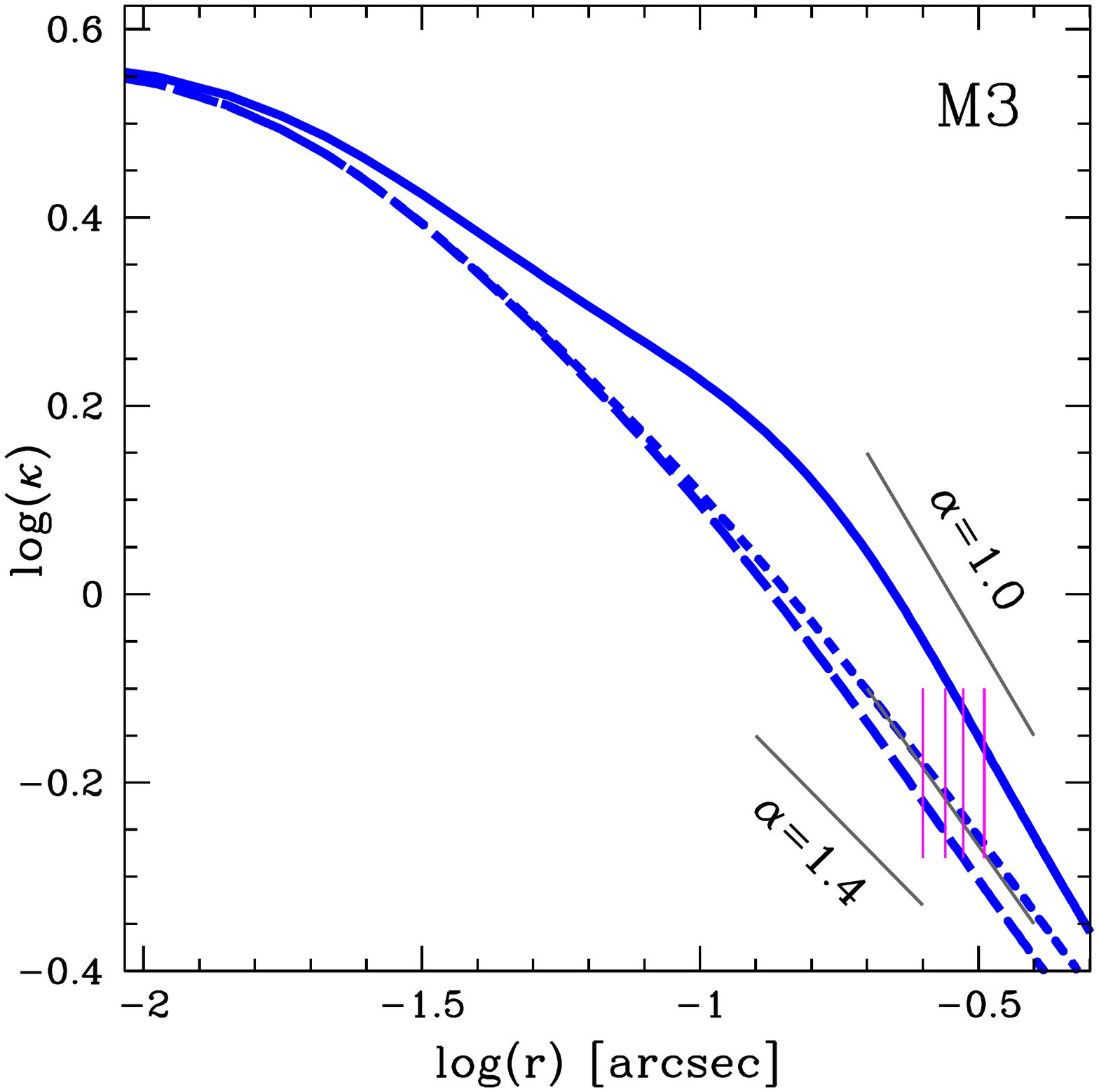}
\vspace{-35pt}
\caption{Radial density profiles of models M1-M3, shown in Figure~\ref{smmagnifs}. Solid and short-dashed blue lines are profiles taken along the positive and negative directions of the major axis, while the long-dashed line is along the positive direction of the axis perpendicular to the major axis. The locations of the 4 images are marked with short vertical magenta lines. Two reference slopes, at $\alpha=1.0$ and $\alpha=1.4$ are also indicated.} 
\label{smprofiles}
\end{figure*} 

\begin{figure*}    %% figure 4
\centering
\vspace{-5pt}
\includegraphics[width=0.32\linewidth]{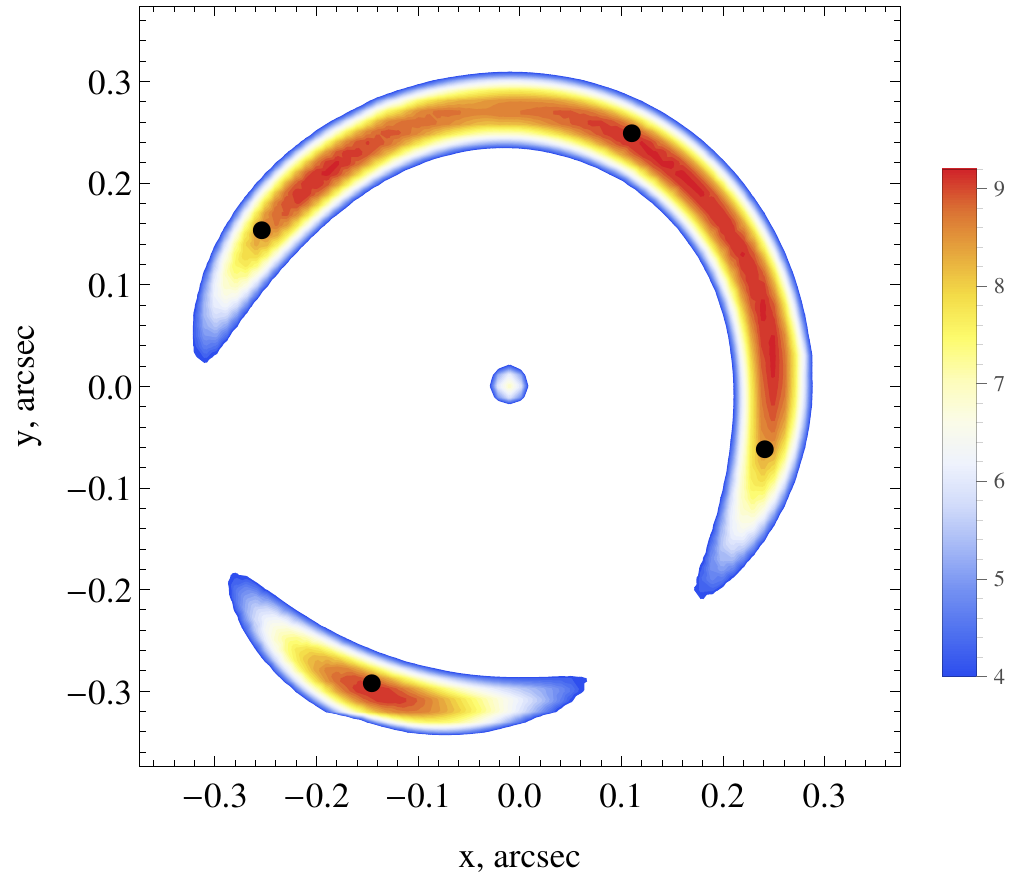}
\includegraphics[width=0.32\linewidth]{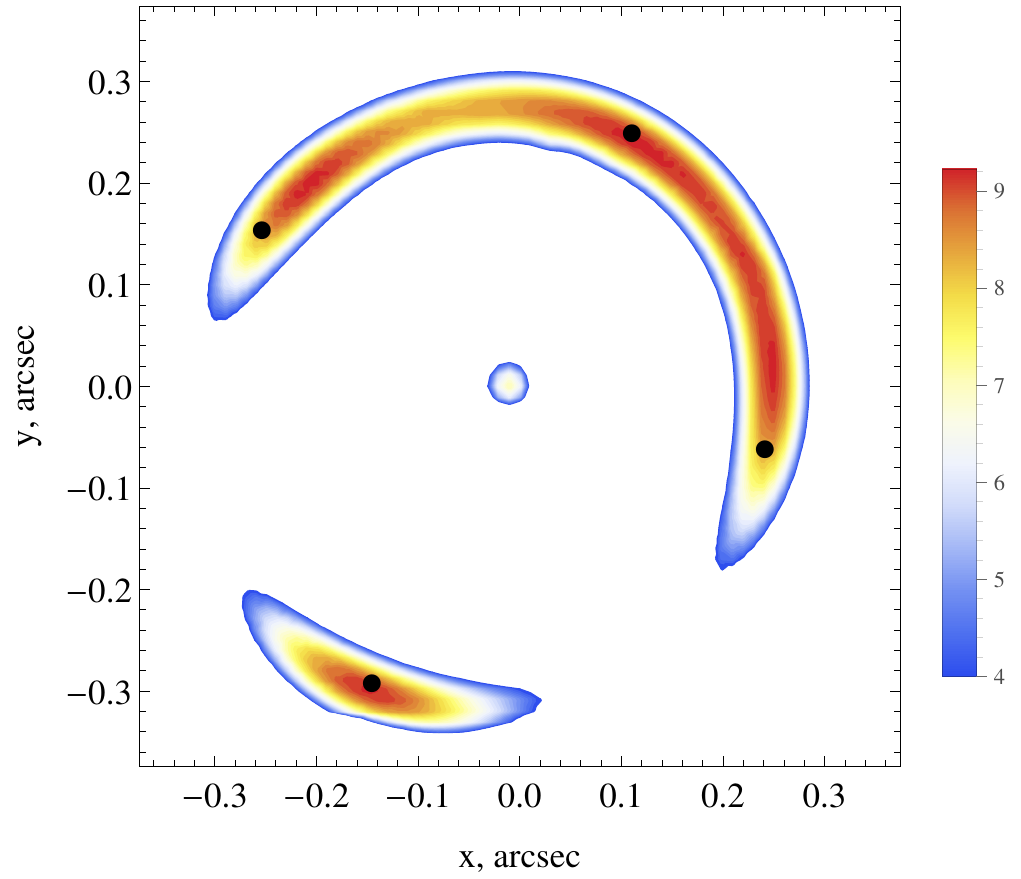}
\includegraphics[width=0.32\linewidth]{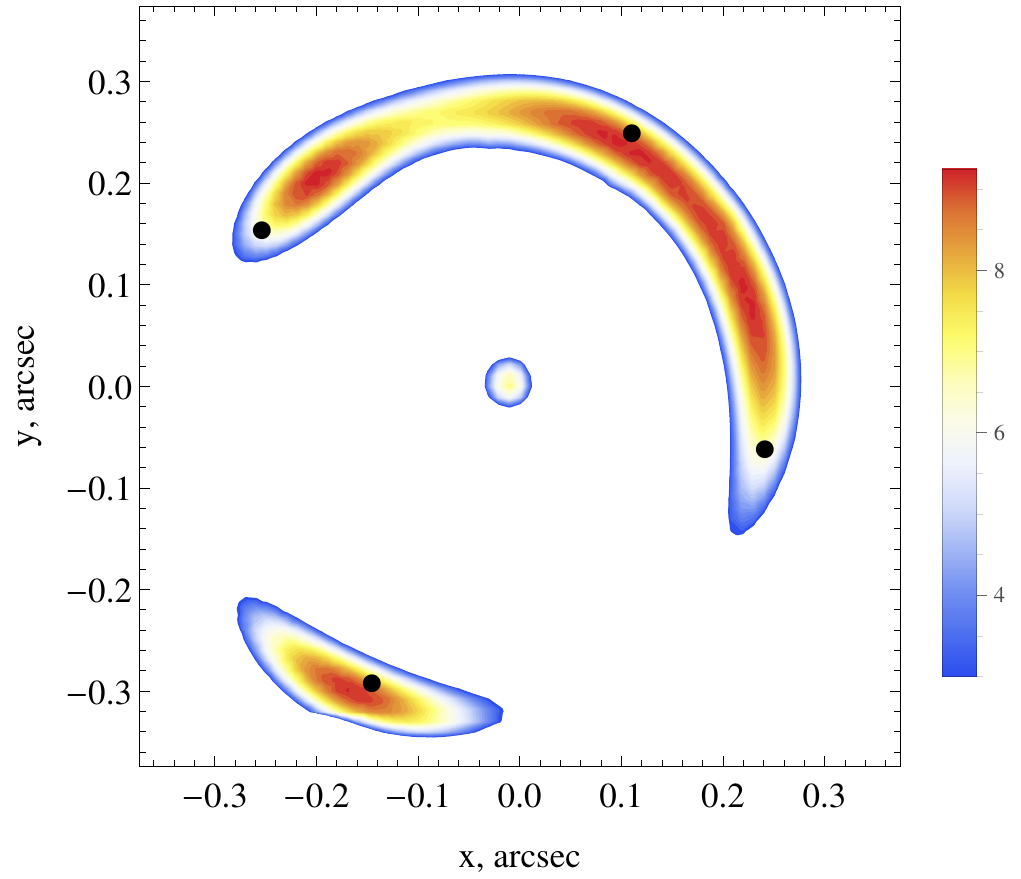}
\caption{Images of extended host galaxy of \sn, lensed by models M1-M3. Images of supernova are shown as black dots, and their IDs are the same as in Figure~\ref{smmassrotate}. The center of the host galaxy and \sn~are displaced by varying amounts, $<0.014''$.} 
\label{smERINGs}
\end{figure*} 

Published works agree on the position angle of the recovered mass distribution, depicted with the thin blue line going through the lens center, pointing at image \#3 to within a few degrees. This elongation direction can be understood even without modeling, from the consideration of image distribution: image \#3 is the 4th arriving, and is usually expected to be closest to the lens center. Since this is not the case here, there has to be mass extention in that direction, to ``push'' the last arriving saddle further from center.  Not surprisingly, our models also show elongation along that axis, though because the total mass distribution is not elliptical, the position angle depends somewhat on the radius.

The axis of the observed light distribution is within the cone outlined by two blue thicker lines, going through the center of the lens in Figure~\ref{smmassrotate}. It is not aligned with the elongation of our mass distributions, echoing the conclusion of  \cite{mor19}. We interpret this misalignment as further indication that the mass, and light, distribution in the lensing galaxy cannot be represented by simple mass models. Given that the reason for the orientation of the mass ellipticity (see the preceding paragraph) is largely model-independent, reconciling it with the position angle of the light could require radial variations in the stellar mass-to-light ratio, or local elongations in the light distribution, such as the ones in the green contours in the southern direction of model M3 in Figure~\ref{smmassrotate}.

The density profiles along three radial spokes emanating from the lens center are shown in Figure~\ref{smprofiles}. The solid and short-dashed blue lines are profiles taken along the positive and negative directions of the major axis, while the long-dashed line is along the positive direction of the axis perpendicular to the major axis. The major axis was determined as the axis of the largest first moment of the total mass distribution. Two reference slopes, of $\alpha=1.0$ and $1.4$ are shown for comparison. Image locations are depicted by vertical magenta line segments. At the location of the images the profile slopes range between $\alpha\sim 1.0$ and $\sim 1.4$, with the average being somewhat steeper than $\alpha=1.3$ slope obtained in \cite{mor19}.

The fact that our mass maps show a variety of isodensity contour shapes means that they are not related by MSD, but instead, by more complex degeneracies that involve transformations of shape \citep{sah06,rea07}, like source plane transformations \citep{sch14}.

\subsection{Microlensing}\label{micro}

Microlensing (de)magnifications implied by our mass models can be deduced from Figure~\ref{smmagnifs}, as the displacement of the magenta point, labeled \sn, from the three large green dots, labeled M1-M3.  One of the features of our models that make them different from the published ones is that we predict solutions whose macro magnifications do not lie on the same diagonal relation on that figure, as those of ellipsoidal power law density profiles. Most interestingly, there is an island of solutions in the right panel of Figure~\ref{smmagnifs} that is closer to \sn~than ellipsoidal power laws; our model M3 is an example. Of the 3 models we consider, changes in flux due to microlensing are smallest for model M3: $\simlt 0.4$ magnitudes for 3 images, and $\sim 0.75$ for the brightest, 1st arriving image. Model M1 would require largest microlensing  contribution, with the 2nd arriving image requiring a demagnification of $\sim 1.1$ magnitudes.

Though our models are different from the elliptical power law models, we agree with the latter that image fluxes cannot be explained without microlensing. Microlensing is almost definitely present in this system, and most likely affects the fluxes of at least three, or all four images. In that sense we agree with the conclusion of \cite{yah17} that microlensing makes it hard to accurately determine the standard candle brightness of the supernova, and therefore use it to help break the mass sheet degeneracy in \sn.

\subsection{Ring image of the extended galaxy host}\label{ring}

Extended images of host galaxies, i.e., rings, have been used extensively in the literature for nearly three decades \citep[e.g.,][]{koc89,koc01}; most recently to help constrain $H_0$ \citep{suy17}, and to uncover mass substructure in the lens galaxy or along the line of sight \citep[e.g.,][]{veg12}. However, just like point images, they are subject to degeneracies \citep{sah01,wal18}, and so may be of limited value in some circumstances. The example in Figure~1 of \cite{den20}, which shows two completely different lens mass distributions producing indistinguishable rings, is a sobering reminder that we do not yet have a full understanding of how lensing degeneracies operate. 

Rings generated from our M1-M3 mass models are shown in Figure~\ref{smERINGs}. The host galaxy source in all 3 cases was assumed to be circular, with a Gaussian light profile. The center of the host galaxy is offset from the supernova by various amounts, $<0.014''$, in these 3 models, comparable to the estimate presented in \cite{mor19}, $\sim 0.02''$. The offset was chosen to better resemble the observed ring, but we did not fine-tune the offset (or other parameters of the source) to produce as a close a match to the ring as possible. Given the appearance of the rings presented in \cite{dha20}, the model that appears closest to observations is M3, because its brightest ring segment is between images $\#1$ and $\#4$. In M1 and M2, on the other hand, the segment between $\#3$ and $\#4$ is equally bright, in contrast to the observations. This suggests that of the 3 models, M3 is the best representation of the lensing galaxy.

\subsection{Time Delays}\label{td}

\begin{figure}    %% figure 5
\centering
\vspace{-5pt}
\includegraphics[width=0.95\linewidth]{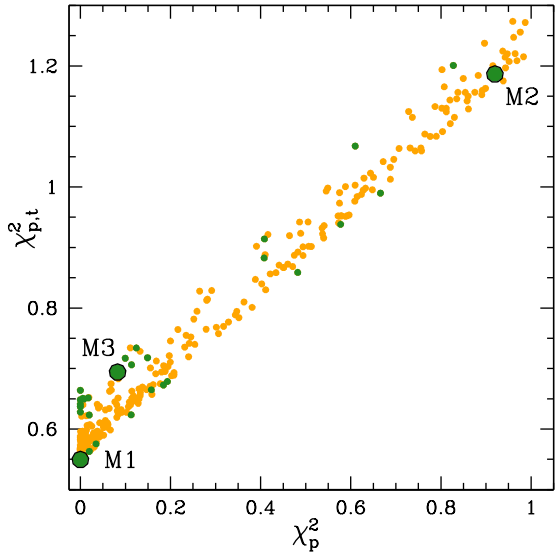}
\vspace{-50pt}
\caption{Distribution of  $\chi^2$ values for our 458 models. The symbols are the same as in Figure~\ref{smmagnifs}. The horizontal and vertical axes contain $\chi^2$ evaluated based on image positions, eq.~\ref{eq:chi2p}, and image positions combined with time delays, eq.~\ref{eq:chi2pt}, respectively. Introducing time delay information does not help constrain models. }
\label{smFOMs}
\end{figure}

We did not consider time delay information when generating our mass models in \S~\ref{modfit}. Had we included observed time delays the model selection would not have been affected. To check that, we calculate $\chi^2$ for all our 458 models, but now using both versions, eq.~\ref{eq:chi2p} and eq.~\ref{eq:chi2pt}. The results are shown in Figure~\ref{smFOMs}.  Because of their large uncertainties, time delays add an approximately constant contribution to each model, and therefore have little discriminating power compared to image positions.

\begin{figure}    %% figure 6
\centering
\vspace{-5pt}
\includegraphics[width=0.98\linewidth]{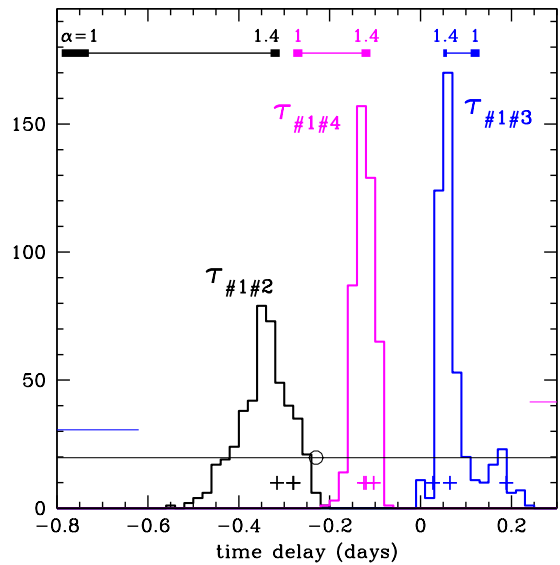}
\vspace{-55pt}
\caption{The three histograms show the distributions of time delays (as labeled) of our models represented by orange points in Figure~\ref{smmagnifs}. The three sets of cross symbols at the bottom show the predictions of the 3 models M1, M2 and M3. The top of the plot shows model predictions of \protect\cite{mor19}. Thick line segments represent $1\sigma$ ranges (their Table 11) for models with potential slopes $\alpha$, as labeled. The thin horizontal lines connect time delays for the whole range of slopes $\alpha= 1\rightarrow 1.4$. The $1\sigma$ uncertainties in the observed time delays are represented by dashed horizontal lines. Note that the central values for $\tau_{\#1\#4}$ and $\tau_{\#1\#3}$ are outside the limits of the plot.} 
\label{smtimedelays}
\end{figure}

Figure~\ref{smtimedelays} shows the distributions of time delays predicted by our lens mass models, orange points in Figure~\ref{smmagnifs}. The three histograms are labeled by the corresponding time delay, using the same image naming convention as in \cite{mor19}. The three sets of three cross symbols show the values from models M1, M2 and M3. The horizontal bars at the top of the plot show \cite{mor19} values, for lens potential profile slopes $\alpha=1$ and $\alpha=1.4$, as labeled. The thin lines show the range of values resulting from the continuous range of $\alpha$'s. While the time delay ranges predicted by our models and these models agree, both are somewhat at odds with observations for $\tau_{\#1\#4}$ and $\tau_{\#1\#3}$, but only at $\simlt 1.5\sigma$, given the uncertainties of $\sim 1$ day. 

\section{Conclusions}

Multiply imaged supernova, and especially Type Ia's, hold a lot of promise for the study of galaxies and cosmology. They are superior to quasars as sources because their time delays are relatively easy to measure, their absolute fluxes are known, and their light eventually fades, allowing observations of the lens galaxy without the glare of the bright images. All these reasons make the very first multiply imaged supernova, \sn, very exciting.

As is often the case, the first of its kind object had some surprizes for modelers. If the mass model is assumed to be an elliptical power law, the mass and light centers are displaced by $\sim 0.1$kpc, which is unlikely to be the case in an equilibrium galaxy. Furthermore, to accommodate microlensing constraints, the power law density slope has to be rather shallow, and so cannot hold at larger radii.

The main motivation for the present paper was to generate lens models whose mass distribution resembles that of local ellipticals, and where the center of observed light coincides with the center of recovered mass.  Similar to local ellipticals whose light is represented by more than one component, our mass models consist of two mass components. The two components are offset, reflecting conclusions of some recent lensing reconstructions. Our final selected models have a single mass peak, and hence coincident mass and light centers. The consequence of using two offset components was to produce dipole-like lopsidedness in the total mass distribution.  Our modeling assumptions, namely two offset elliptical potentials, result in a range of mass models---not a single model---all satisfying positional and time delay image information.  For illustration purposes we concentrated on three of these models, based on their magnification properties, but could have chosen a different set.

Though our mass distributions are not elliptical power laws, we do agree with the published models on some key features of the lens:
(1) To explain the fluxes of the \sn, microlensing by stars in the lens is required for at least 3, or all 4 images of the quad; (2) The total mass density profile at the location of the images is significantly shallower than isothermal; (3) The position angle of galaxy mass elongation is misaligned with the light, which is probably an indication that the mass, and light, distribution in the lens is not simple.

We note that our two-component models may not be the only way to account for the observations of \sn. It is possible that the main lensing galaxy has a simple one-component structure, but there are additional significant secondary galaxies, either at the same redshift or along the line of sight. However, there are no observed nearby galaxies, making this scenario less likely. The line-of-sight structure (LoS) is also unlikely because the source is very nearby, $z=0.409$, so the line of sight dimension is short.  \cite{des18} estimate that the typical number of LoS subhalos along a given direction is much less than 1, and their mass is a small fraction of the mass enclosed by the images.

\section*{Acknowledgments}
The authors would like to thank Edvard M\"ortsell for kindly providing addition information about their modeling, and the referee for helping us improve the paper. This work was supported by the NSF through grant \#1757388.

%\bibliographystyle{mnras}
%\bibliography{bibfile} 

\end{document}